\documentclass[%
 reprint,
 longbibliography,
superscriptaddress,
 amsmath,amssymb,
 aps,
prb,
]{revtex4-2}

\usepackage{graphicx}
\usepackage{tikz}
\usepackage{dcolumn}
\usepackage{bm}

\usepackage[utf8]{inputenc}
\usepackage[T1]{fontenc}
\usepackage{mathptmx}
\usepackage{siunitx}
\usepackage{amsmath}
\usepackage{lipsum}
\usepackage{array}
\usepackage[normalem]{ulem}
\usepackage{xcolor}
\usepackage[ colorlinks,
    linkcolor={blue},
    citecolor={blue},
    urlcolor={blue}]{hyperref}
\usepackage{braket}

\newcommand{\reviewout}[1]{}

\newcommand\eqt{\hspace{0.17em}{=}\hspace{0.17em}}

\newcommand\get{\hspace{0.17em}{\ge}\hspace{0.17em}}

\newcommand\mt{\hspace{0.17em}{-}\hspace{0.17em}}

\newcommand{\br}{\mathbf{r}}

\newcommand{\JWout}[1]{}
\newcommand{\XGout}[1]{}

\newcommand{\rl}{\br_\ell}
\newcommand{\rlp}{\br_{\ell'}}

\begin{document}

\preprint{APS/123-QED}

\newcommand{\papertitle}{Optical absorption spectra from the \textit{GW}-Bethe-Salpeter framework for thousands of atoms}
\title{\papertitle}

\author{Maximilian Graml}
\email{maximilian.graml@physik.uni-regensburg.de}
\affiliation{\mbox{Institute of Theoretical Physics and Regensburg Center for Ultrafast Nanoscopy (RUN),} \mbox{University of Regensburg, 93053 Regensburg, Germany}}
\author{Ritaj Tyagi}
\email{ritaj.tyagi@physik.uni-regensburg.de}
\affiliation{\mbox{Institute of Theoretical Physics and Regensburg Center for Ultrafast Nanoscopy (RUN),} \mbox{University of Regensburg, 93053 Regensburg, Germany}}
\author{R\'emi Pasquier}
\affiliation{\mbox{Institute of Theoretical Physics and Regensburg Center for Ultrafast Nanoscopy (RUN),} \mbox{University of Regensburg, 93053 Regensburg, Germany}}
\author{\v{S}t\v{e}p\'{a}n Marek}
\affiliation{\mbox{Institute of Theoretical Physics and Regensburg Center for Ultrafast Nanoscopy (RUN),} \mbox{University of Regensburg, 93053 Regensburg, Germany}}
\affiliation{Halle-Berlin-Regensburg Cluster of Excellence CCE, University of Regensburg, 93053 Regensburg, Germany}
\author{Jan Wilhelm}
\email{jan.wilhelm@physik.uni-regensburg.de}
\affiliation{\mbox{Institute of Theoretical Physics and Regensburg Center for Ultrafast Nanoscopy (RUN),} \mbox{University of Regensburg, 93053 Regensburg, Germany}}
\affiliation{Halle-Berlin-Regensburg Cluster of Excellence CCE, University of Regensburg, 93053 Regensburg, Germany}

\date{\today}

\begin{abstract}
The frequency-dependent absorption and emission of light determine the colors of matter and underpin today's technologies, from solar cells to light-emitting diodes. Predicting optical properties from first principles remains a computational challenge, especially for large and disordered systems. The $GW$ plus Bethe-Salpeter equation ($GW$-BSE) approach provides predictive accuracy for optical properties, but conventional implementations become costly as system size grows. Here, we present a $GW$-BSE framework combining real-time propagation, atom-centered basis sets, and a locally optimized real-space resolution of the identity. The cost of each propagation step scales quadratically with system size and roughly equals the cost of one self-consistent-field iteration in a density-functional calculation with a semilocal functional. This computational efficiency allows  us to calculate the absorption spectrum of a nanographene containing 4616 atoms and opens a path toward routine calculations of optical properties for large disordered systems.
\end{abstract}

\maketitle

Interactions between light and matter lie at the heart of   physics, chemistry and biology. When matter absorbs or emits light, its electrons interact with the electromagnetic field. The resulting optical absorption and emission spectra are closely connected to the atomic structure and chemical environment of the system and optical spectra can therefore reveal the arrangement of atoms and molecules. A quantitative understanding of  these phenomena requires a microscopic description of the electronic excitations involved:  electrons are promoted to higher-energy states and interact with the positively charged hole they leave behind. Electron and hole are quasiparticles, bare charges dressed by the screening response of all other electrons.

Calculating these excitations becomes challenging when the atomic structure required to describe them involves a large number of atoms. In liquids and solutions, calculated optical  properties must be converged with respect to the size of the simulation cell and the number of solvent molecules. At finite temperature, an additional requirement is statistical sampling: spectra must be calculated for several structures drawn from a molecular-dynamics trajectory and then averaged. Calculations of optical absorption spectra of liquid water and ice illustrate both requirements~\cite{Nguyen2019,Tang2025}. Large porous and nanostructured materials present the same system-size challenge. Examples include excitons in metal-organic frameworks and electronic states in moir\'e structures of twisted two-dimensional materials~\cite{Kshirsagar2021,Graml2024,Naik2022}. The computational effort therefore increases with both the number of atoms in each structure and, when statistical sampling is required, the ensemble size. This creates a need for methods to compute optical properties that remain computationally efficient for large  systems.

Density-functional theory (DFT) provides a framework for determining ground-state properties from the electron density~\cite{HohenbergKohn1964,KohnSham1965,Martin2020}. Light absorption, however, probes the response of the electrons to a time-dependent electromagnetic field, which drives the electrons out of the ground state. Time-dependent density-functional theory (TDDFT) extends the density-functional framework to this response and is often used to calculate electronic excitations and optical spectra~\cite{RungeGross1984,Ullrich2011}. With commonly used local or semilocal approximations to the exchange-correlation kernel, TDDFT does not   capture the long-range electron-hole attraction required for bound excitons   in condensed matter  and for charge-transfer excitations in molecules and molecular compounds~\cite{Onida2002,Reining2002, Peach2008,Ullrich2011, Loos2021}.

The established many-body approach to these excitations combines the $GW$ approximation with the Bethe-Salpeter equation (BSE)~\cite{Hedin1965,SalpeterBethe1951,Albrecht1998,Rohlfing1998,Rohlfing1999,Rohlfing2000,Onida2002,Sander2015, Ljungberg2015,Blase2020,Vorwerk2019,Baumeier2012,Graml2026}.  The $GW$ calculation determines the energies for adding or removing an electron (quasiparticle energies) and hence the fundamental band gap.  The BSE then describes neutral excitations by coupling electron-hole transitions  through the screened direct electron-hole attraction and the bare electron-hole exchange. BSE provides excitation energies and amplitudes from which exciton binding energies, the spatial structure of excitons, and optical spectra can be obtained~\cite{Albrecht1998,Rohlfing1998,Rohlfing1999,Rohlfing2000,Onida2002, Blase2011, Baumeier2012,Sander2015,Liu2020, Bruneval2016,Vorwerk2019,Blase2020,Gonze2020,Zhou2025,Graml2026}.

In a conventional BSE calculation, the excitonic Hamiltonian contains  combinations of occupied and virtual electronic states, so its dimension grows as the product of the numbers of these states~\cite{Rohlfing1998,Rohlfing2000,Onida2002,Vorwerk2019}. Direct diagonalization of this Hamiltonian yields the excitation energies and amplitudes, and hence the properties described above, but storing and diagonalizing the matrix soon becomes computationally demanding as the system grows. Several approaches reduce this cost by avoiding the full diagonalization and representing the electron-hole interaction efficiently~\cite{Ljungberg2015,Marsili2017,Krause2017,Henneke2020,FoersterVisscher2022,Gao2022,Yu2024,Maurer2026}. 
An alternative approach is real-time BSE (RT-BSE), where the electronic response is propagated in real time after perturbation by a short electric-field pulse~\cite{Schmidt2003,Attaccalite2011,Perfetto2015,Rabani2015,Chan2021,Marek2025,Tumakov2026,Merkel2024}. A Fourier transform of the time-dependent polarization gives the optical absorption spectrum; its peak positions yield excitation energies and its spectral weights yield oscillator strengths. 
Real-time propagation offers the possibility of substantially increased computational efficiency~\cite{Schmidt2003,Rabani2015,Merkel2024}. 
It turns out that evaluating the time steps in RT-BSE is particularly efficient in a space-local representation of the involved quantities. 
Plane-wave calculations can obtain localized orbitals by transforming their extended electronic states into Wannier functions~\cite{Merkel2024}. 
Atom-centered basis functions, in contrast, are local by construction and require no such transformation.

We consider the BSE in the Tamm-Dancoff approximation (TDA), which neglects   the coupling between excitations and de-excitations~\footnote{Real-time propagation without the TDA~\cite{Attaccalite2011,Marek2025} is also implemented in CP2K~\cite{CP2KDevelopmentTrunk} but is not considered here.}.  The equation of motion follows from the Kadanoff-Baym equations for nonequilibrium Green's functions by approximating the self-energy with the time-local screened exchange~\cite{Attaccalite2011,Martin2016,Marek2025}.
In the TDA, only the occupied-virtual block $\varrho_{ia}(t)\eqt\langle\psi_i|\Delta\hat\rho(t)|\psi_a\rangle$ of the induced density matrix $\Delta\hat\rho(t)$ is propagated, where $\psi_i/\psi_a$ denotes an occupied/virtual molecular orbital. For the closed-shell systems considered here, the spin-independent field drives both spin channels identically, and $\varrho_{ia}(t)$ refers to one spin channel.

At the beginning of the time propagation, a weak $\delta$-electric-field kick $\mathbf F$ is applied which gives for the occupied-virtual block $\varrho_{ia}$ in linear response to the field~$\mathbf{F}$
\begin{align}
\varrho_{ia}(0^+)
 =  -iq\,\mathbf{F}\cdot\mathbf r_{ia}/\hbar\,,\hspace{1.5em}
\mathbf r_{ia}
&={\int} d\mathbf r\,\psi_i(\mathbf r)\,\mathbf r\,\psi_a(\mathbf r)\,,
\label{eq:initial_density}
\end{align}
where $q$ is the electronic charge, $\mathbf r_{ia}$ is the transition dipole and $\mathbf F\eqt-\int dt\,\mathbf E(t)$ specifies the pulse strength and direction. 
The  density matrix $\varrho_{ia}(t)$ is then propagated  via
\begin{align}
\partial_t\varrho_{ia}(t)
&=\frac{i}{\hbar}\left[
(\varepsilon_a^{GW}-\varepsilon_i^{GW})\varrho_{ia}(t)
+V^{\mathrm H}_{ia}(t)
+\Sigma^{\mathrm{SX}}_{ia}(t)
\right]\,. 
\label{eq:rtbse_tda}
\end{align}
Here, $\varepsilon_i^{GW}$ and $\varepsilon_a^{GW}$ are occupied and virtual quasiparticle energies obtained from a single-shot $GW$ calculation ($G_0W_0$). Equation~\eqref{eq:rtbse_tda} resembles the von Neumann equation~\cite{Ullrich2011} for independent-particle dynamics: the term containing the quasiparticle energies describes the independent-particle time evolution, whereas $V^{\mathrm H}$ and $\Sigma^{\mathrm{SX}}$ account for electron-electron interaction effects. The Hartree term $V^{\mathrm H}$ describes the electron-hole exchange interaction, and $\Sigma^{\mathrm{SX}}$ describes the screened direct Coulomb interaction between the electron and hole.
With $\varrho_{ia}$ per spin channel, Eq.~\eqref{eq:rtbse_tda} is the real-time counterpart of the singlet BSE in the appendix of Ref.~\cite{Graml2026}: the Hartree term is built from the spin-summed induced density, which gives the factor 2 of the bare exchange term in the singlet kernel [Eq.~\eqref{eq:hartree_ao_si}], whereas the screened exchange acts within one spin channel. The triplet channel is not excited by the field.

We propagate Eq.~\eqref{eq:rtbse_tda} with the fourth-order Runge-Kutta solver~\cite{CRecipesPressFlannery}, which requires to compute $V^{\mathrm H}_{ia}$ and $\Sigma^{\mathrm{SX}}_{ia}$ from the density matrix. 
To do so, we transform the  density matrix~$\varrho_{ia}$ to the atomic-orbital basis and the real-space grid~$\{\rl\}$,
\begin{align}
\varrho(\rl,\rlp;t)
&=\sum_{\mu\nu}\phi_\mu(\rl)\,
\Big[\sum_{ia}C_{\mu i}\,\varrho_{ia}(t)\,C_{\nu a}\Big]
\,\phi_\nu(\rlp)\,.
\label{eq:grid_density}
\end{align}
The grid~$\{\rl\}$ arises from the real-space resolution of the identity (RI-RS), also known as separable density fitting, which represents orbital products through their values at selected points~\cite{Parrish2013,LuYing2015,Duchemin2019,Duchemin2021,Delesma2024}. Its construction is discussed below. The coefficients $C_{\mu i}$ and $C_{\nu a}$ expand the occupied and virtual molecular orbitals in atomic orbitals (AOs): $\psi_p(\mathbf r)\eqt\sum_\mu C_{\mu p}\phi_\mu(\mathbf r)$.

The screened-exchange self-energy is evaluated as
\begin{align}
&\Sigma^{\mathrm{SX}}(\rl,\rlp;t)
=-\varrho(\rl,\rlp;t)\,W(\rl,\rlp;\omega=0)\,,
\label{eq:sx_grid}\\[0.5em]
&\Sigma^{\mathrm{SX}}_{ia}(t)
=\sum_{\mu\nu}C_{\mu i} \,\Big[
\sum_{\ell\ell'}\phi_\mu(\mathbf r_\ell)
\,\Sigma^{\mathrm{SX}}(\rl,\rlp;t)
\,\phi_\nu(\mathbf r_{\ell'})
\Big]\,C_{\nu a}\,.
\label{eq:sx_mo}
\end{align}
The screened Coulomb interaction $W(\rl,\rlp;\omega=0)$ is computed once for the ground state and is not updated during RT-BSE propagation.   The Hartree term in Eq.~\eqref{eq:rtbse_tda} is computed as described in Sec.~\ref{sec:si_interactions} of the Supplementary Information (SI).

For computing the optical absorption spectrum, we consider a pulse along Cartesian direction~$n$ with time-integrated electric field $F_{n}$. From the propagated  density matrix, the induced polarization in linear order in $F_n$ along direction~$m$ is
\begin{align}
P_m^{(n)}(t)
&=2q\,\operatorname{Tr}\!\left[\hat r_m\,\Delta\hat\rho^{(n)}(t)\right]
=4q\,\operatorname{Re}\sum_{ia}r_{m,ai}\,\varrho_{ia}^{(n)}(t)\,.
\label{eq:polarization_time}
\end{align}
Here, $r_{m,ai}\eqt (\br_{ia})_m$; the factor 2 sums the dipole over the two spin channels, the counterpart of the singlet normalization of the transition dipole in Ref.~\cite{Graml2026}, and the second equality uses $\varrho_{ai}\eqt\varrho_{ia}^{*}$. After Fourier transformation and Pad\'e interpolation~\cite{Kick2024,Leucke2025}, we obtain the polarizability:
\begin{align}
\alpha_{mn}(\omega)={P_m^{(n)}(\omega)}/{F_n}\,.
\label{eq:directional_absorption}
\end{align}

\begin{figure}[t!]
  \centering
  \begin{tikzpicture}
    \node[inner sep=0] (plot) {\includegraphics[width=\columnwidth]{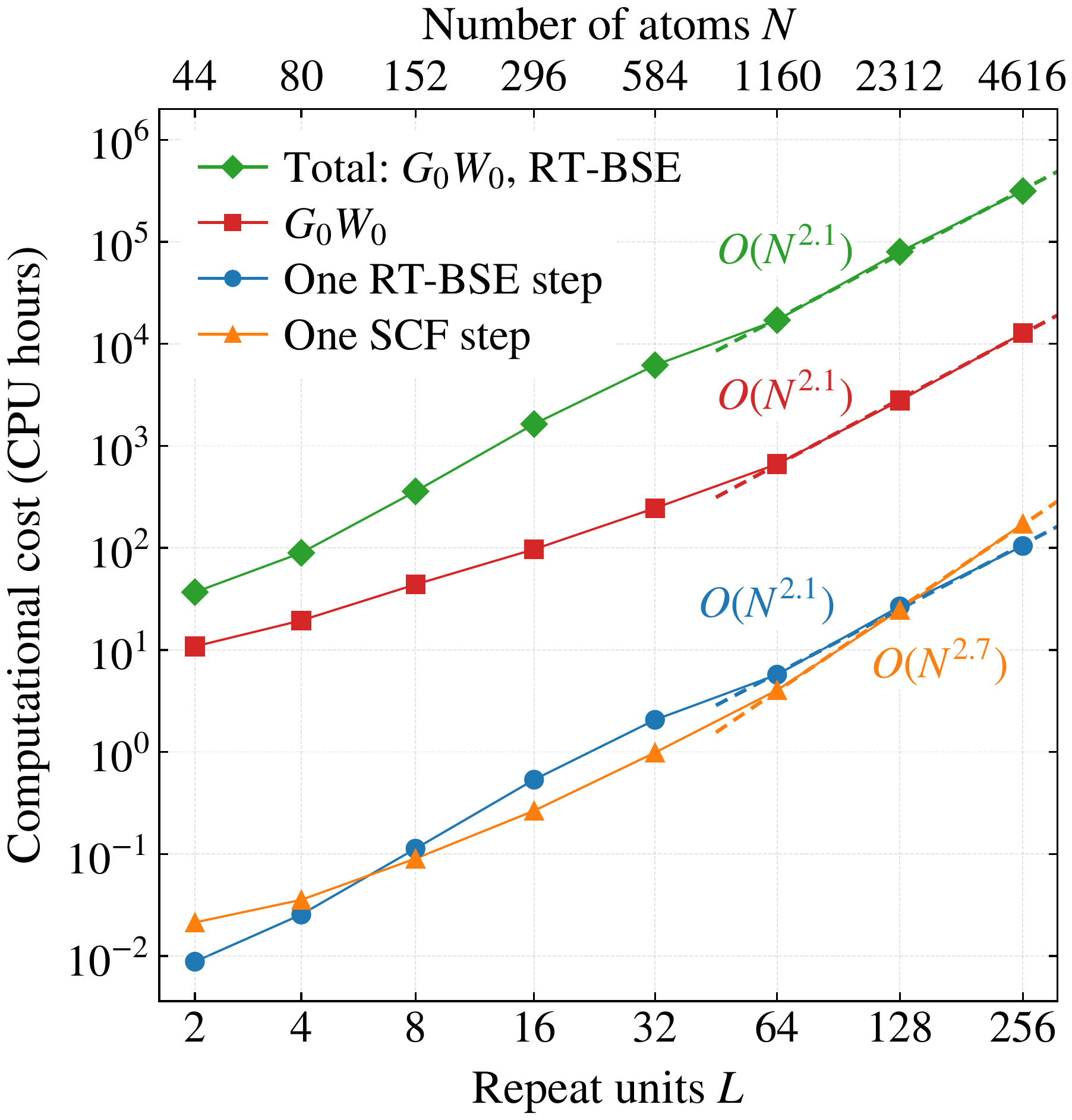}};
    \begin{scope}[shift={(plot.north east)},xshift=-2.457cm,yshift=-6.722cm,scale=0.85]
      \input{figure/fig1_ribbon_geometry.tex}
    \end{scope}
  \end{tikzpicture}
  \caption{Computational cost of calculating the optical absorption spectra of the nanographenes shown in the inset from RT-BSE (Eqs.~\eqref{eq:initial_density}--\eqref{eq:directional_absorption}; absorption spectra in Fig.~\ref{fig:length-sweep}). Dashed lines fit the three largest nanographenes and fitted exponents are displayed. Computational details are given in the SI, Sec.~\ref{sec:compdetails}.}
  \label{fig:timing}
\end{figure}

\begin{figure*}[t]
  \centering
  \includegraphics[width=\textwidth]{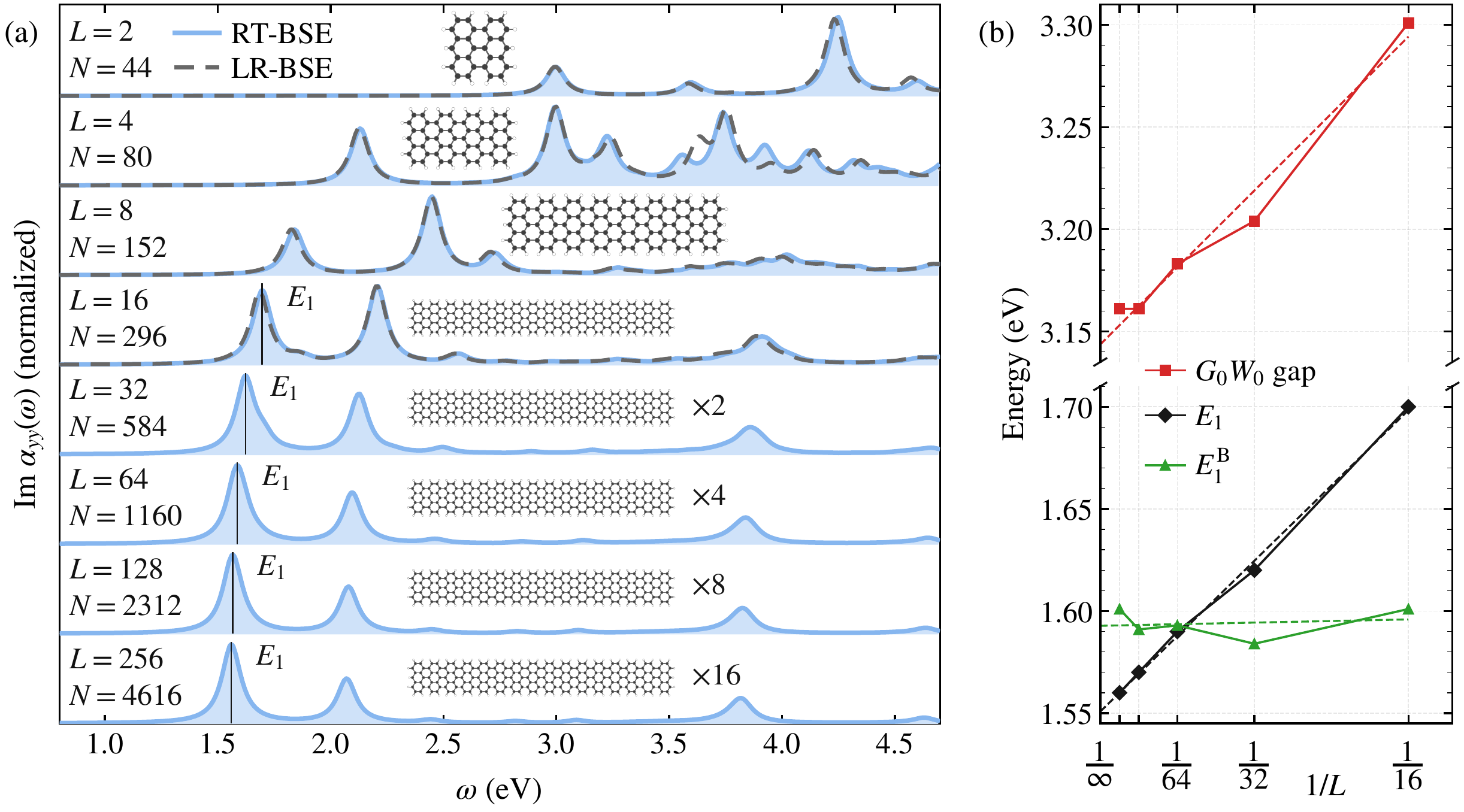}
  \caption{Optical properties of nanographenes. (a) Absorption spectra computed from real-time propagation (RT-BSE, solid blue; Eqs.~\eqref{eq:initial_density}--\eqref{eq:directional_absorption}) and, for $L\eqt2,4,8,16$, from diagonalization of the BSE excitonic Hamiltonian (LR-BSE, dashed grey). (b) $G_0W_0$ gap, first excitation energy $E_1$ indicated in panel (a), and positive binding energy $E_1^{\mathrm B}\eqt E_{\mathrm{gap}}^{GW}\mt E_1$ as functions of inverse nanographene length $1/L$.}
  \label{fig:length-sweep}
\end{figure*}

We directly turn to discussing the computational cost of the algorithm in Eqs.~\eqref{eq:initial_density}\,--\,\eqref{eq:directional_absorption}, benchmarked on the nanographenes shown in the inset of Fig.~\ref{fig:timing}.  Each time step involves matrix-matrix multiplications (Eqs.~\eqref{eq:grid_density} and \eqref{eq:sx_mo}) and a Hadamard product (Eq.~\eqref{eq:sx_grid}) on the grid. As shown in Fig.~\ref{fig:timing}, the computational effort of a single RT-BSE time step (blue) is of the same order of magnitude as one DFT self-consistent-field (SCF) iteration (orange; PBE functional~\cite{PerdewBurkeErnzerhof1996}). The key to bringing the RT-BSE step cost down to that of an SCF iteration is that the number of real-space grid points $\{\rl\}$ is of the same order as the number of AOs, because SCF matrix operations take place in the AO basis. To the best of our knowledge, previous RI-RS $GW$ calculations used pre-tabulated atomic grids~\cite{Duchemin2019,Duchemin2021,Delesma2024}. Here, we first select 20\% of those points using an interpoint-distance criterion (convergence test for the number of selected grid points in the SI, Fig.~\ref{fig:params}d). We then optimize the selected points to match RI-RS three-center integrals to analytical values. This optimization is carried out on an atomic cluster for each atom carved out from the nanoribbon structure, comprising all atoms within a 3-\AA\ sphere around it (convergence test in the SI, Fig.~\ref{fig:params}a); details will be given in a forthcoming work~\cite{Tyagi2026rirs,Tyagi2026}. For this optimization, we use 27 grid points per H atom and 67 per C atom. To exploit spatial locality, we set $\phi_\mu(\rl)$ to zero when grid point $\rl$ lies more than 7~\AA\ from the atom on which orbital $\mu$ is centered (convergence test in the SI, Fig.~\ref{fig:params}b). For the screened interaction in Eq.~\eqref{eq:sx_grid}, we drop grid pairs $(\rl,\rlp)$ with $|\rl-\rlp|>50~\text{\AA}$ (convergence test in the SI, Fig.~\ref{fig:params}c). This cutoff is motivated by the exciton size:  $(\rl,\rlp)$ pairs farther apart than 50~\AA\ lie beyond the spatial extent of the exciton (SI, Fig.~\ref{fig:exciton-size}).   For $L\ge16$, where the nanographene length exceeds 50~\AA, the cutoff makes    $\Sigma^\text{SX}$ sparse, consistent with the quadratic scaling of one RT-BSE step observed in Fig.~\ref{fig:timing}. 
The green curve in Fig.~\ref{fig:timing} gives the measured cost of the complete run, including ground-state DFT calculation restarted from a converged wavefunction, $G_0W_0$, and the RT-BSE real-time propagation ($\sim$\,2800 time steps). The $G_0W_0$ calculation is  2.3 ($L\eqt 2$) to  27 times ($L\eqt 128$) cheaper than the  RT-BSE propagation.

The optical absorption spectra computed from RT-BSE are shown in Fig.~\ref{fig:length-sweep}a.  For $L\eqt2,4,8,16$, we compare them with linear-response BSE (LR-BSE) spectra obtained under TDA by diagonalizing the excitonic Hamiltonian in the occupied-virtual transition space~\cite{Graml2026}.  The spectra agree almost perfectly: the first absorption peak differs by no more than about 0.01~eV, and the next two low-energy peaks are closely reproduced. The clearest discrepancy occurs for $L\eqt4$ between 3.5 and 4~eV, where two closely spaced peaks merge in the real-time spectrum; we presume this is caused by the analytic continuation in $G_0W_0$.
For $L\ge32$, diagonalization becomes too expensive, so we use RT-BSE to follow the absorption spectrum up to $L\eqt256$ ($N\eqt18L+8\eqt4616$ atoms). The first excitation energy $E_1$ approaches the 1D-periodic   limit as the nanographene grows~\footnote{Note that when using BSE@ev$GW_0$@PBE, the extrapolated 1D-periodic limit of the lowest excitation energy is 2.1 eV~\cite{Graml2026} which is aligning well with the experimental value~\cite{Denk2014}. Here, we instead use $G_0W_0$ to avoid numerical instabilities arising in low-scaling ev$GW_0$ that will be addressed in an upcoming work.}. Figure~\ref{fig:length-sweep}b shows that $E_1$ decreases as $1/L$,  following the $1/L$ dependence of the $G_0W_0$ gap.  This behavior is consistent with the confinement of the quasiparticle states along the finite nanographene. Their difference, the positive exciton binding energy $E_1^{\mathrm B}$, is nearly independent of length for $L\ge16$. This agrees with the saturation of the exciton size  for $L\get 16$~\cite{Graml2026} (SI, Fig.~\ref{fig:exciton-size}).  

We conclude with an outlook towards excited-state calculations in extended systems. Existing $GW$-BSE studies of liquids and moir\'e structures have required substantial computational effort~\cite{Tang2025,Graml2024,Naik2022}. With periodic boundary conditions, the presented RT-BSE algorithm could make routine $GW$-BSE calculations for liquids and solids, including solid-liquid interfaces and moir\'e structures, feasible across larger cells, including statistical sampling. The compact real-space grids could also benefit time propagation based on the Kadanoff-Baym equations, providing a route to dynamical screening beyond the static BSE kernel used here~\cite{Perfetto2015,SchroedterBonitz2024}. A further prospect is excited-state dynamics with moving nuclei~\cite{Jiang2021}, for example within an Ehrenfest approach~\cite{PerfettoStefanucci2023}. The $GW$ calculation may then become a dominant cost because quasiparticle energies and screening are updated at each new nuclear configuration, whereas each BSE propagation step has approximately the cost of a single SCF iteration (Fig.~\ref{fig:timing}).

\section*{Data Availability}
The atomic geometries, input and output files, tables, and plotting scripts are available on GitHub~\cite{TyagiRTBSENanoribbonData}.  The implementation is in the current CP2K development trunk~\cite{CP2KDevelopmentTrunk,Kuhne2020CP2K,Iannuzzi2026CP2K,Wilhelm2026CP2KDynamics}.

\begin{acknowledgments}
We thank F.~A.~Delesma and D.~Golze for helpful discussions.
J.W.~acknowledges the DFG for funding via the Emmy Noether Programme (project number 503985532),  RTG 2905 (project number 502572516), and CRC 1277 (project number 314695032).
This work was supported by the German Research Foundation (DFG) through Germany's Excellence Strategy--EXC 3112/1--533767171 (Center for Chiral Electronics).
The authors gratefully acknowledge the computing time provided to them on the high-performance computers Noctua 2 and Otus~\cite{Bauer2024Noctua2,Ehtesabi2025Otus} at the NHR Center PC2. These systems are funded by the Federal Ministry of Education and Research and the state governments participating on the basis of the resolutions of the GWK for the national high-performance computing at universities (www.nhr-verein.de/unsere-partner).
\end{acknowledgments}

\nocite{Denk2014}
\bibliography{main}

\clearpage
\onecolumngrid
\begin{center}
{\large\bfseries Supporting Information to\par}
\vspace{0.6em}
{\large\bfseries \papertitle\par}
\vspace{1.2em}
{\normalsize Maximilian Graml$^{1}$, Ritaj Tyagi$^{1}$, R\'emi Pasquier$^{1}$, \v{S}t\v{e}p\'{a}n Marek$^{1,2}$, and Jan Wilhelm$^{1,2}$\par}
\vspace{0.5em}
{\normalsize\itshape $^{1}$Institute of Theoretical Physics and Regensburg Center for Ultrafast Nanoscopy (RUN),\par}
{\normalsize\itshape University of Regensburg, 93053 Regensburg, Germany\par}
{\normalsize\itshape $^{2}$Halle-Berlin-Regensburg Cluster of Excellence CCE, University of Regensburg, 93053 Regensburg, Germany\par}
\end{center}

\setcounter{equation}{0}
\renewcommand{\theequation}{S\arabic{equation}}
\renewcommand{\theHequation}{S\arabic{equation}}
\setcounter{figure}{0}
\renewcommand{\thefigure}{S\arabic{figure}}
\renewcommand{\theHfigure}{S\arabic{figure}}
\setcounter{section}{0}
\renewcommand{\thesection}{S\arabic{section}}
\renewcommand{\theHsection}{S\arabic{section}}

\section{Real-space interaction kernels and Hartree contribution}
\label{sec:si_interactions}

We describe how the Coulomb potentials are represented on the real-space RI-RS grid. We calculate the static screened Coulomb interaction $W(\omega\eqt0)$ in the auxiliary RI basis $\{\varphi_P\}$ in the $GW$ calculation and transform it using the RI-RS coefficients $Z_{\ell P}$~\cite{Duchemin2019,Duchemin2021,Delesma2024,Tyagi2026rirs,Tyagi2026}. The bare Coulomb interaction is transformed analogously:
\begin{align}
W(\rl,\rlp;\omega=0)
&=\sum_{PQ}Z_{\ell P}\,W_{PQ}^{(\omega=0)}\,Z_{\ell'Q}
&&\text{(for $\Sigma^{\mathrm{SX}}$)}\,,
\label{eq:w_grid_si}\\
V(\rl,\rlp)
&=\sum_{PQ}Z_{\ell P}\,V_{PQ}\,Z_{\ell'Q}
&&\text{(for $V^{\mathrm H}$)}\,,
\label{eq:v_grid_si}
\end{align}
The Hartree contribution is computed from
\begin{align}
V^{\mathrm H}_{\mu\nu}(t)
&=2\sum_{\ell\ell'}\phi_\mu(\mathbf r_\ell)\,\phi_\nu(\mathbf r_\ell)\,
V(\rl,\rlp)\,\varrho(\rlp,\rlp;t)\,,
\label{eq:hartree_ao_si}\\
V^{\mathrm H}_{ia}(t)
&=\sum_{\mu\nu}C_{\mu i}\,V^{\mathrm H}_{\mu\nu}(t)\,C_{\nu a}\,.
\label{eq:hartree_mo_si}
\end{align}
The factor 2 in Eq.~\eqref{eq:hartree_ao_si} is the spin sum of the induced density, i.e.\ the factor 2 of the bare exchange term in the singlet BSE kernel~\cite{Graml2026} (main text, after Eq.~\eqref{eq:rtbse_tda}). 
The screened-exchange term, Eq.~\eqref{eq:sx_grid}, acts within one spin channel and carries no such factor.

\section{Computational details}\label{sec:compdetails}
\begin{figure*}[p]
  \centering
  \includegraphics[width=\textwidth]{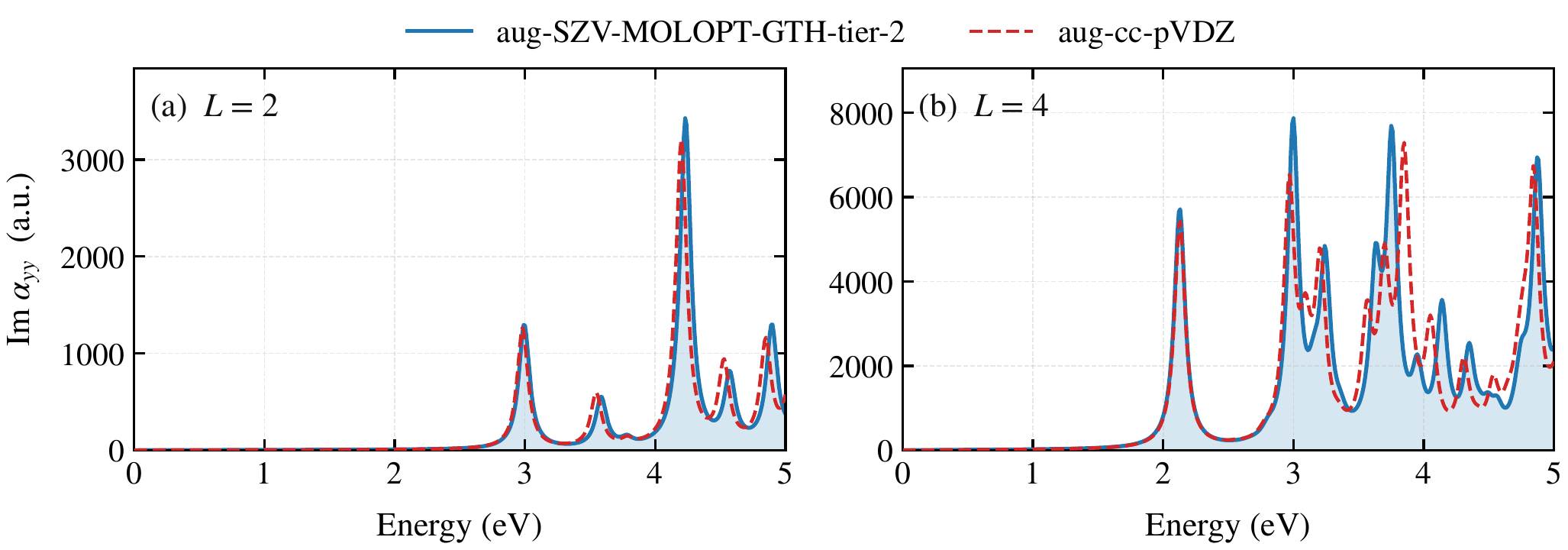}
  \caption{Basis-set convergence. LR-BSE spectra of the (a)~$L=2$ and (b)~$L=4$ ribbons in the aug-SZV-MOLOPT-GTH-tier-2 basis used throughout (solid, shaded) and in aug-cc-pVDZ (dashed).}
  \label{fig:basis}
  \vspace{5.0em}
  \includegraphics[width=\textwidth]{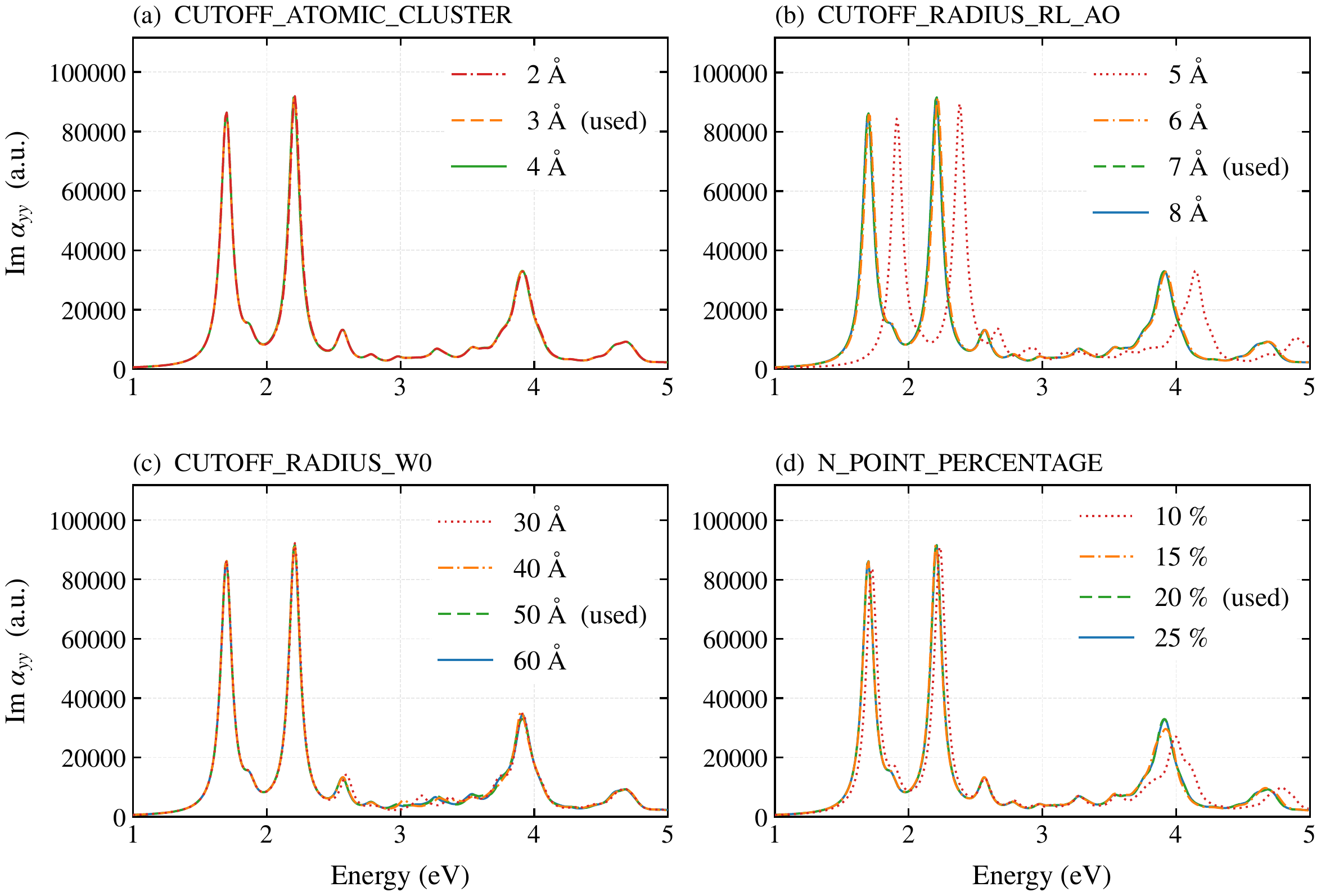}
  \caption{Numerical convergence tests for the $L=16$ optical absorption spectrum: (a) atomic-cluster radius for grid optimization, (b) real-space AO product cutoff, (c) screened-interaction cutoff, and (d) fraction of initial grid points used. Production settings are marked ``used''.}
  \label{fig:params}
\end{figure*}

The absorption spectra of the nanoribbons were calculated in the Tamm-Dancoff approximation from $G_0W_0$ quasiparticle energies starting with Perdew-Burke-Ernzerhof (PBE) orbitals and eigenvalues. The calculations used the aug-SZV-MOLOPT-GTH-tier-2 basis~\cite{WilhelmAugMoloptBasis}; basis-set convergence was checked against all-electron aug-cc-pVDZ~\cite{Kendall1992} (Fig.~\ref{fig:basis}). Imaginary-time and imaginary-frequency integrations used 30 minimax quadrature points~\cite{Azizi2023,Azizi2024}.  The Gaussian-and-plane-wave density cutoff was 600~Ry. Propagation lasted 200~fs with an exponential damping time of 13.2~fs, corresponding to a 0.05-eV broadening; the direct BSE spectra use the same Lorentzian width. The four RI-RS and propagation parameter checks are in Fig.~\ref{fig:params}. Each grid is optimized on a 3-\AA\ atomic cluster including neighbors; 20\% of the original grid points optimized for the separable RI-RS approach~\cite{Duchemin2019,Duchemin2021} are used, namely 27 per H and 67 per C atom.

The propagation length of 200~fs corresponds to a direct Fourier frequency spacing $2\pi\hbar/T\eqt0.0207$~eV; the spectra are sampled on a 0.01-eV energy grid after Pad\'e continuation~\cite{Leucke2025}. Occupied and virtual orbitals $\psi_i$ and $\psi_a$ used in the time-propagation equation, Eq.~\eqref{eq:rtbse_tda}, are constrained to lie within 10~eV of their respective Kohn-Sham band edges.   The maximum included quasiparticle energy difference $\varepsilon_a^{GW}-\varepsilon_i^{GW}$ fixes the fourth-order Runge-Kutta stability limit. For $L\eqt16$, the maximum quasiparticle energy difference is 26.6~eV, giving a 69.9-as time step and 2860 time steps.

\begin{figure*}[p]
  \centering
  \includegraphics[width=0.58\textwidth]{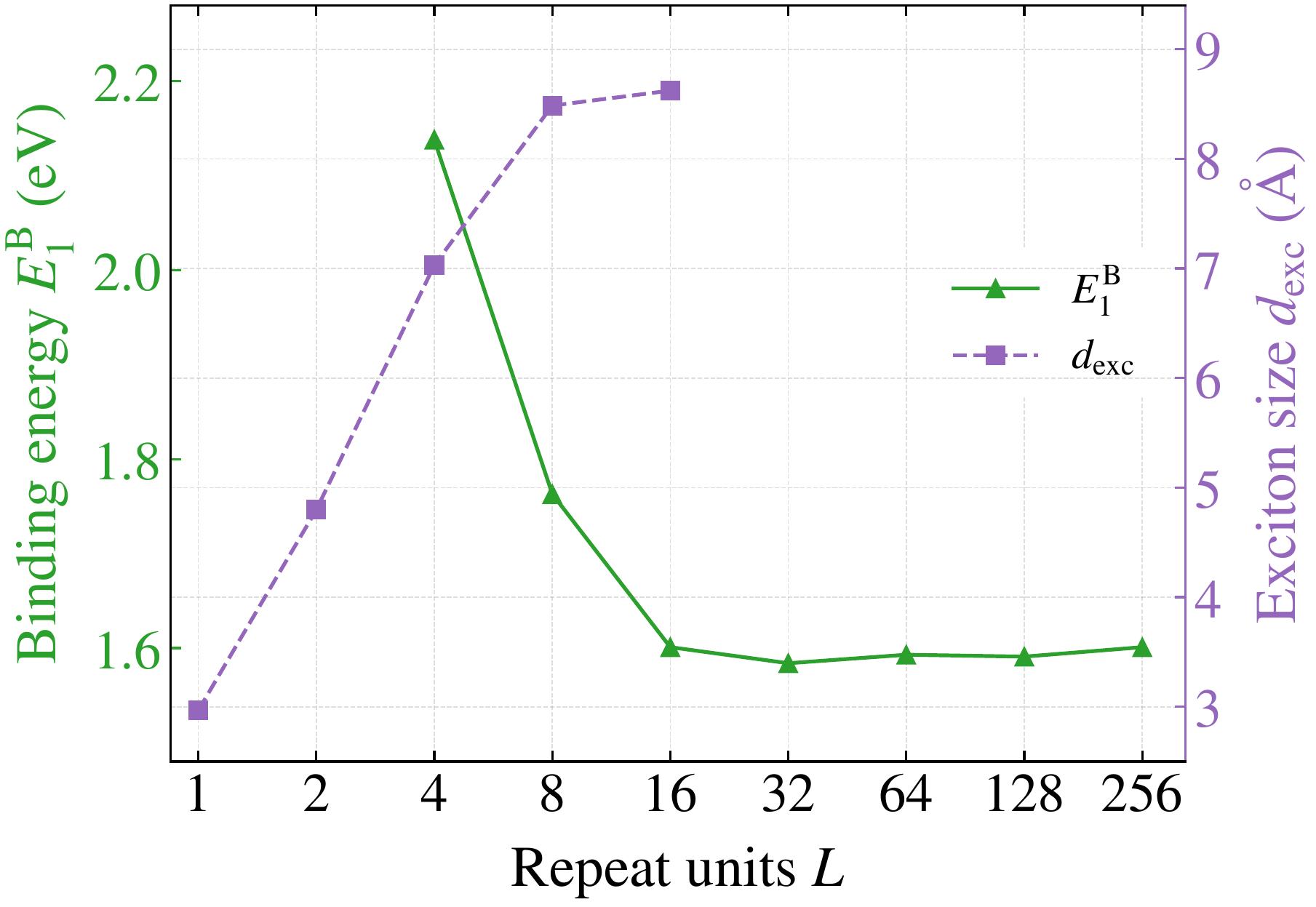}
  \caption{Binding energy $E_1^{\mathrm B}\eqt  E_{\mathrm{gap}}^{GW}\mt E_1$ and exciton size  $d_{\mathrm{exc}}$~\cite{Graml2026} versus nanographene length.}
  \label{fig:exciton-size}
\end{figure*}

\end{document}